\documentclass[aps,prd,twocolumn,superscriptaddress,nofootinbib,10pt]{revtex4-1}

\usepackage[utf8]{inputenc}
\usepackage{graphicx,amsmath,amsfonts,amssymb,aas_macros,slashed}
\usepackage{bbold}
\usepackage{datetime}
\usepackage{numprint}
\usepackage{enumitem}
\usepackage{graphicx,color,amsmath}
\usepackage{hyperref}

\usepackage[normalem]{ulem}

\usepackage{rotating}
\allowdisplaybreaks

\usepackage[normalem]{ulem} 

\usepackage{pgfplots}
\usepackage{bm}
\usepackage{mathtools}

\makeatletter
\newcommand{\thickhline}{%
    \noalign {\ifnum 0=`}\fi \hrule height 1pt
    \futurelet \reserved@a \@xhline
}
\newcolumntype{"}{@{\hskip\tabcolsep\vrule width 1pt\hskip\tabcolsep}}
\makeatother

\begin{document}

\title{Atmospheric axion-like particles at Super-Kamiokande}

\author{Kingman Cheung}
\email{cheung@phys.nthu.edu.tw}
\affiliation{Department of Physics, National Tsing Hua University,	Hsinchu 300, Taiwan}
\affiliation{Center for Theory and Computation, National Tsing Hua University,	Hsinchu 300, Taiwan}
\affiliation{Division of Quantum Phases and Devices, School of Physics, Konkuk University, Seoul 143-701, Republic of Korea}

\author{Jui-Lin Kuo}
\email{juilink1@uci.edu}
\affiliation{Department of Physics and Astronomy,
University of California, Irvine, CA 92697-4575, USA}

\author{Po-Yan Tseng}
\email{tpoyan1209@gmail.com}
\affiliation{Department of Physics, National Tsing Hua University,	Hsinchu 300, Taiwan}
\affiliation{Center for Theory and Computation, National Tsing Hua University,	Hsinchu 300, Taiwan}

\author{Zeren Simon Wang}
\email{wzs@mx.nthu.edu.tw}
\affiliation{Department of Physics, National Tsing Hua University,	Hsinchu 300, Taiwan}
\affiliation{Center for Theory and Computation, National Tsing Hua University,	Hsinchu 300, Taiwan}

\begin{abstract}

We consider a muonphilic axion-like-particle (ALP), denoted as $a$, lighter than twice the muon mass.
ALPs of this mass range dominantly decay into a pair of photons, induced by a triangular muon loop.
Such light muonphilic ALPs are naturally long-lived.
At the atmosphere, the ALPs are copiously produced from charged-meson decays in air showers, such as $\pi^\pm \to \mu^\pm \nu a$, via the ALP-muon coupling $g_{a\mu\mu}$.
After propagating tens of kilometers, the ALPs decay with $a\to \gamma \gamma$ inside large-volume Cherenkov detectors near the Earth's surface, such as Super-Kamiokande (SK).
We find the present SK observation constrains on muonphilic ALPs of mass range [1 MeV, 30 MeV] and ALP-muon coupling $[10^{-3}$, $10^{2}]$, assuming the proper decay length $c\tau_a$ in [$10^{-3}$ km, $10^6$ km] either dependent on or independent of $g_{a\mu\mu}$.
We conclude that atmospheric searches of such exotic states can be complementary to collider and beam-dump experiments as well as astrophysical probes.

\end{abstract}

    \maketitle

\section{Introduction}

The strong CP problem~\cite{Peccei:1977hh,Peccei:1977ur,Weinberg:1977ma,Wilczek:1977pj} in the Standard Model (SM) can be solved by introducing a global $U(1)_{\text{PQ}}$ symmetry which was spontaneously broken down by a dynamical CP-conserving axion field.
The corresponding pseudo-Nambu-Goldstone-boson of the broken symmetry is called the QCD axion, which in addition serves as a dark matter candidate~\cite{Preskill:1982cy,Abbott:1982af,Dine:1982ah}.
The breaking scale of the new symmetry should be high: $f_a\gtrsim 10^9$ GeV~\cite{Feng:1997tn}, demanding tiny masses of the QCD axion and their couplings to the SM particles, since the latter two are inversely proportional to $f_a$.
This results in a very long lifetime of the QCD axions.

A closely-related hypothetical particle is known as axion-like particle (ALP), which, like the QCD axion, is also a pseudoscalar boson.
Unlike the QCD axion, the ALP mass is not linearly proportional to the couplings to the SM particles, and the ALP hence does not necessarily fix the strong CP problem.
However, the ALP remains one of the possible dark matter candidates,  and its mass could possibly range across more than 20 orders of magnitude~\cite{Kim:2015yna,DeMartino:2017qsa,Rubakov:1997vp}.
Further, such ALPs appear in various theoretical models beyond the SM~\cite{Svrcek:2006yi,Gelmini:1980re,Davidson:1981zd,Wilczek:1982rv}.

In general, the ALPs can couple to photons, leptons, quarks, as well as gauge bosons at either tree level or loop level.
The phenomenology with only ALP-photon interactions $g_{a\gamma\gamma}$ has been vastly investigated (see Ref.~\cite{Agrawal:2021dbo} and the references therein).
In particular, for sub-GeV ALPs, PRIMEX~\cite{Aloni:2019ruo} and Belle II~\cite{Belle-II:2020jti} provide the most stringent upper bounds on $|g_{a\gamma\gamma}|$, and ALPs of mass $m_a\lesssim 30$  MeV~\cite{Agrawal:2021dbo,Buen-Abad:2021fwq} are disfavored by beam-dump experiments.
However, for ALP-muon interactions, only BaBar~\cite{BaBar:2016sci} gives constraints, for ALPs heavier than twice the muon mass~\cite{Buen-Abad:2021fwq}. 
As far as we know, muonphilic ALPs lighter than twice the muon mass have not been directly constrained.
Therefore, we choose to focus on this scenario in the present work.

When cosmic rays reach the Earth's atmosphere, large atmospheric air showers are produced including copious production of pseudoscalar mesons.
Such mesons can decay to light long-lived particles (LLPs) (see Refs.~\cite{Curtin:2018mvb,Lee:2018pag,Alimena:2019zri} for reviews on LLPs), which travel macroscopic distances before decaying potentially in the large-volume neutrino experiments at the Earth's surface.
This allows to probe various models predicting such LLPs including heavy neutral leptons~\cite{Kusenko:2004qc,Asaka:2012hc,Masip:2014xna,Arguelles:2019ziu,Coloma:2019htx,Meighen-Berger:2020eun}, the lightest neutralinos in the R-parity-violating supersymmetry~\cite{Candia:2021bsl}, light dark matter~\cite{Arguelles:2022fqq,Su:2020zny}, axion dark radiation~\cite{Gu:2021lni,Cui:2022owf}, and milli-charged particles~\cite{ArguellesDelgado:2021lek}.
Similarly, the muonphilic ALPs can be abundantly produced via 
charged-meson decays from the atmosphere air showers.
Such ALPs should be long-lived, because both they are very light and their decay channels are radiatively suppressed if their mass is below twice the muon mass.
After traveling tens of kilometers across the atmosphere, these ALPs may subsequently decay into two photons inside the detectors of neutrino experiments, such as Super-Kamiokande (SK).
With the tool \texttt{MCEq}~\cite{Fedynitch:2015zma}, we numerically compute the ALPs' flux from the air showers including the propagation through dense medium.
We then estimate the signal event rates at the SK detector, which is sensitive to events of energy below $\mathcal{O}(100)~{\rm GeV}$~\cite{Super-Kamiokande:2017yvm}.
After discussing the background events, we obtain SK bounds on both physical observables and model parameters.

This article is organized as follows.
In Sec.~\ref{sec:model}, we introduce the theoretical scenario we investigate in this work.
The estimation of the ALP flux from the air showers is detailed in Sec.~\ref{sec:alp_flux}, followed by Sec.~\ref{sec:alp_detection} and Sec.~\ref{sec:SuperK} explaining the ALP detection on the Earth and introducing the SK experiment, respectively.
The final numerical results are presented and discussed in Sec.~\ref{sec:results}.
At the end, Sec.~\ref{sec:conclusions} provides a summary and outlook of this work.

\section{ALP-muon interactions}\label{sec:model}

\begin{figure*}[t]
\begin{center}
\includegraphics[width=0.48\textwidth]{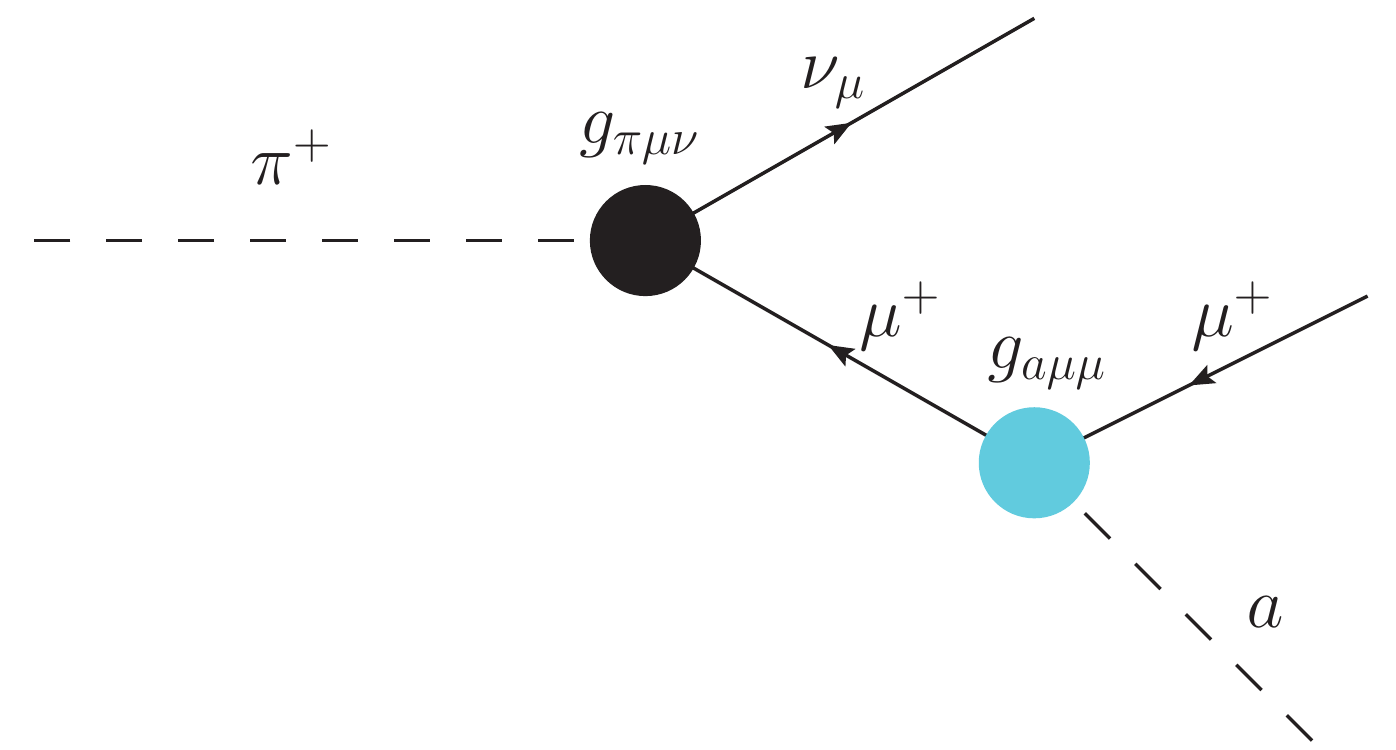}\hfill
\includegraphics[width=0.48\textwidth]{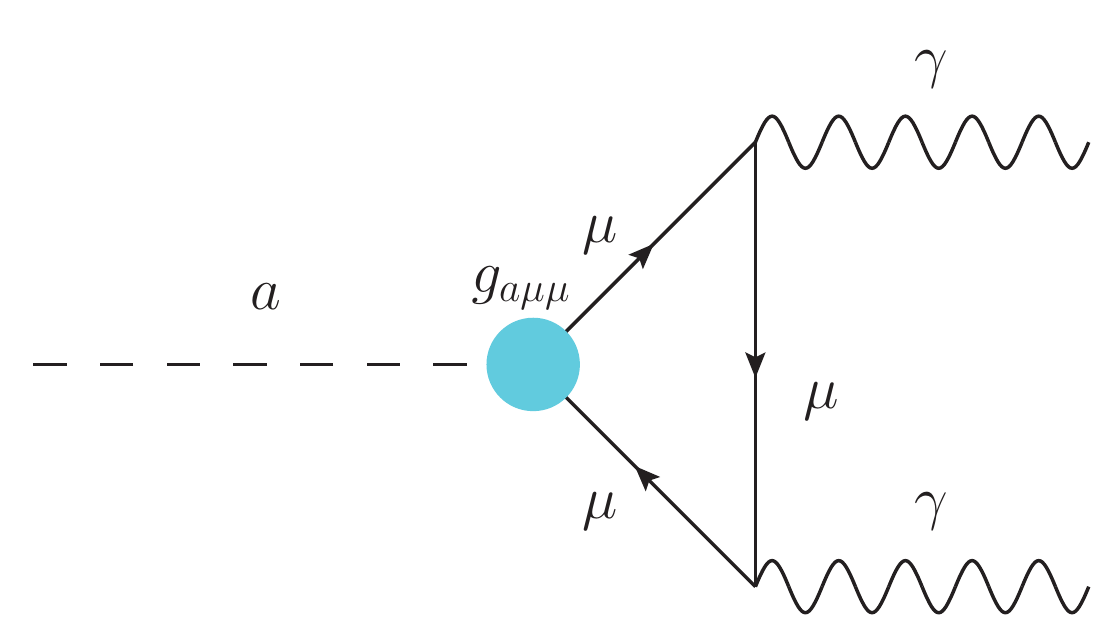}
\end{center}
\caption{\textit{Left panel}: Feynman diagram for production of the muonphilic ALP from the charged-pion decay. 
The decay vertex factor $g_{\pi \mu\nu}$ denotes the effective 
coupling for the charged pion decay $\pi^+ \to \mu^+ \nu_\mu$.
Its conjugated diagram with $\pi^-$ decays is not shown here.
\textit{Right panel}: Feynman diagram for the decay of the muonphilic ALP into a pair of photons, via a triangular muon loop.
}
\label{fig:feynman}
\end{figure*}

In this work, we consider the interactions between the ALP and muon with an effective Lagrangian expressed as 
\begin{align}
\mathcal{L} \supset -i g_{a\mu\mu} a \bar{\mu} \gamma_5 \mu\,,
\label{eq:ALP-muon_Lagrangian}
\end{align}
where $g_{a\mu\mu}$ is a dimensionless coupling constant.
For ALP mass $m_a$ larger than twice the muon mass $m_\mu$, the ALP can decay into a pair of muons, while for a lighter ALP only decays into a SM photon pair induced by a triangular muon loop. 
The loop-induced interaction between ALP and photons can be described by  
\begin{align}
\mathcal{L}_{\rm loop} \supset -\dfrac{1}{4} g_{a\gamma\gamma}^{\rm eff} a F^{\mu\nu} \tilde{F}_{\mu\nu}\,,
\end{align}
with the effective coupling being~\cite{Bauer:2017ris}
\begin{align}
g_{a\gamma\gamma}^{\rm eff} = \dfrac{g_{a\mu\mu} \alpha}{m_\mu \pi } \left[1- \dfrac{4m_\mu^2}{m_a^2} \arcsin^2 \left(\dfrac{m_a}{2m_\mu} \right)\right]\,,
\end{align}
which is valid for $m_a \le 2 m_\mu$.
The lifetime of ALP with $m_a < 2m_\mu$ then reads
\begin{align}
\label{eq:ALP_lifetime}
 \tau_a = \Gamma_{a\rightarrow \gamma \gamma}^{-1} = \dfrac{64\pi}{(g_{a\gamma\gamma}^{\rm eff})^{2} m_a^{3}}\,.
\end{align}

With the ALP-muon coupling, ALPs can be produced from charged-meson decays in air showers, dominated by the decay of charged pions $\pi^\pm \rightarrow \mu^\pm \nu a$; therefore, the kinematically allowed ALP mass range is $0 \leq m_a \leq m_\pi - m_\mu$ assuming massless SM neutrinos.

In Fig.~\ref{fig:feynman}, we present two Feynman diagrams for the production and decay of the ALPs, respectively.

The ALP-muon and ALP-photon couplings both contribute to the muon magnetic dipole moment, $a_\mu=(g-2)_\mu/2$~\cite{Buen-Abad:2021fwq}.
The one-loop result of $g_{a\mu\mu}$ leads to negative contributions to $a_\mu$.
However, if we also include the ALP-photon coupling, $g_{a\gamma\gamma}$, the two couplings will induce two-loop light-by-light and Barr-Zee diagrams~\cite{Chun:2015xfx,Marciano:2016yhf}, which, in combination, provide positive contributions to $a_\mu$~\cite{Keung:2021rps}.
On the experimental side, the updated combined results of Fermilab~\cite{Muong-2:2021ojo} and BNL~\cite{Muong-2:2006rrc} measurements indicate a 4.25$\sigma$ positive deviation from the SM theoretical prediction:
\begin{eqnarray}
\Delta a^{\rm BNL}_\mu &=& a^{\rm BNL}_\mu-a^{\rm SM}_\mu = (251\pm 59) \times 10^{-11}\,.
\end{eqnarray}
However, theoretical uncertainties arising from hadronic vacuum polarization may alleviate the tension between these measurements and the SM~\cite{Borsanyi:2020mff}.
Given the large uncertainties within the SM computation, we do not take into account $(g-2)_\mu$ in our analysis.

\section{ALP flux from air showers}\label{sec:alp_flux}

We utilize the numerical code \texttt{MCEq}~\cite{Fedynitch:2015zma} to compute the ALP flux at the Earth's surface. 
\texttt{MCEq} numerically solves cascade equations of particles propagating in a dense medium; in this work, we use it to study the ALP production throughout the cascade of secondary cosmic rays.
We adopt the \texttt{H3a} parameterization of the cosmic ray flux at the top of atmosphere provided in Ref.~\cite{Gaisser:2011klf} and take the hadronic interaction model \texttt{SIBYLL2.3c} in Ref.~\cite{Fedynitch:2018cbl}.
The atmosphere is modeled by the CORSIKA parameterizations of the U.S. Standard Atmosphere~\cite{Heck:1998vt}.

To implement the process $\pi^\pm \rightarrow \mu^\pm \nu a$ in \texttt{MCEq}, we compute the corresponding decay matrix
\begin{align}
D^{ij}_{\pi^\pm \rightarrow a} = \Delta T_{\pi^\pm}^i \dfrac{dN_a}{dT_a} (T_{\pi^\pm}^i, T_a^j)\,,
\end{align}
where $T_{\pi^\pm}$ and $T_a$ are the kinetic energy of the pion and 
the ALP in the lab frame, with $i,j$ and $\Delta T_{\pi^\pm}^i$ being the kinetic energy bin indices and width.
%
The ALP energy spectrum $dN_a/dT_a$ in the lab frame is obtained by applying a Lorentz boost to the energy spectrum in the pion rest frame
\begin{align}
\label{eq:energy_spectrum_lab}
\dfrac{dN_a}{dT_a} = \int \dfrac{d\Omega}{4\pi} \dfrac{dN_a}{dE_a^*} \left| \dfrac{\partial E_a^*}{\partial T_a}\right|\,,
\end{align}
where $E_a^*$ is the energy of the ALP in the pion rest frame and $|\partial E_a^*  /\partial T_a |$ is the Jacobian between $E_a^*$ and $T_a$.
Detailed derivation of $dN_a/dT_a$ is given in App.~\ref{app:energy_spectrum}\footnote{The $dN_a/dT_a$ distribution is found to be smooth enough so that the bin width implemented by \texttt{MCEq} should have minimal aliasing effect.}.

After the decay matrix is tabulated, we augment the decay channels of $\pi^\pm$ with $\pi^\pm \rightarrow \mu \nu a$. 
We first consider the case that the ALP decay into two photons is unrelated to the production coupling $g_{a\mu\mu}$, \textit{i.e.}, the ALP flux at production is proportional to $g_{a\mu\mu}^2$ while the decay is determined by the decay length $c\tau_a$ in the ALP rest frame. 
Note that we assume that the overall distribution of other decay products such as muon and neutrino is not affected by the newly added channel as its branching ratio is suppressed.
The results can be easily reinterpreted for other theoretical scenarios where the atmospheric charged pions decay to an LLP which then subsequently decays visibly in the SK detector.
Then, we study the case that both production and decay depend on the coupling constant $g_{a\mu\mu}$. 

\section{ALP detection on the Earth}\label{sec:alp_detection}

After arriving at the Earth, the ALP can decay into two photons through a muon loop, with a lifetime given in Eq.~\eqref{eq:ALP_lifetime}.
The photons so-produced can then be detected by the Cherenkov detector in neutrino experiments. 
Given zenith angle $\theta$, detector geometry, and data-taking time $\Delta t$, the event distribution can be calculated by
\begin{align}
\dfrac{d^2 N_{\rm event}}{dT_ad\cos\theta} = \epsilon  \Delta t A_{\rm eff}(T_a,\cos\theta) \dfrac{d^2 \Phi_a}{dT_ad\cos\theta}\,,
\end{align}
where $\epsilon$ is the detection efficiency and we use the output of $\texttt{MCEq}$ for the differential flux $d^2 \Phi_a/(dT_ad\cos\theta)$.
The computation of effective detection area $A_{\rm eff}$, depending on $T_a$ and $\cos\theta$, is given in the appendix of Ref.~\cite{Arguelles:2019ziu} and is also demonstrated in App.~\ref{app:Aeff} for completeness. 
The main SM background of such a two-photon signal from the ALP decay 
consists of neutral pions decaying into two photons, and neutrino-induced electron-like events that create multiple Cherenkov rings in the electromagnetic showers. 
In 5326 live days, the number of these events have been studied in Ref.~\cite{Super-Kamiokande:2017yvm}, with the best-fit values being 1727 and 797, respectively.
These background events will be taken into account in Sec.~\ref{sec:results} when we estimate the sensitivity reach of SK.

In addition to the signals from ALP decaying into a $\gamma$-pair, ALP can interact with atoms in the detector to create mono-$\gamma$ signal with an energy similar to the energy of ALP, the so-called inverse-Primakoff process.
The cross section of inverse-Primakoff process was studied in details in \cite{Abe:2020mcs}, which can be expressed as 
\begin{align}
\sigma_{\rm IP} \simeq \left(\dfrac{g_{a\gamma\gamma}^{\rm eff}}{1\,{\rm GeV}^{-1}}\right)^2 \times 2\,{\rm GeV}^{-2}.
\end{align}
However, since $A_{\rm eff}$ for detecting the ALP decay is larger by orders of magnitude than the effective cross section of inverse-Primakoff process $N_T \sigma_{\rm IP}$ with $N_T$ being the total number of target atoms inside the fiducial volume of the detector, we can infer that the event rate from the ALP decay dominates over that from the inverse-Primakoff process; therefore, we will not consider this possibility further in this work.

\section{Super-Kamiokande}
\label{sec:SuperK}

We note that only when the charged pions have a kinetic energy below their critical energy\footnote{As the branching ratio of $\pi^\pm \to \mu^\pm \nu a$ is suppressed, we adopt the critical energy of the charged pion predicted in the SM.} $\epsilon_{\pi^\pm} = 115\,{\rm GeV}$~\cite{Gaisser:2016uoy} and hence a small Lorentz boost, do they essentially all decay well before reaching the Earth's surface; therefore, ALP flux at $T_a \gtrsim \epsilon_{\pi^\pm}$ is strongly suppressed.
In order to maximize the sensitivity, we focus on the water-based Cherenkov detector of Super-Kamiokande (SK), which has good energy resolution in the sub- and multi-GeV ranges~\cite{Super-Kamiokande:2017yvm}.

Following the analysis in Ref.~\cite{Super-Kamiokande:2017yvm}, the geometry of the SK detector is assumed to be a cylinder with radius $R_{\rm SK} = 20\,{\rm m}$ and height $H_{\rm SK} = 40\,{\rm m}$. The lifetime of SK is taken to be 5326 days with a flat detection efficiency of $0.75$.
Fully-contained events in SK can be grouped into different categories according to the energy and configuration of observed Cherenkov rings.
Since the signal from ALP decay constitutes two electron-like Cherenkov rings, we consider data of $\pi^0$-like two-ring events in 5 energy-bins for sub-GeV $T_a$ and electron-like multi-ring events in 5 $\cos\theta$-bins for multi-GeV $T_a$ provided in Ref.~\cite{Super-Kamiokande:2017yvm}.

\section{Constraints on the parameter space}\label{sec:results}

\begin{figure*}[t]
\begin{center}
\includegraphics[width=0.48\textwidth]{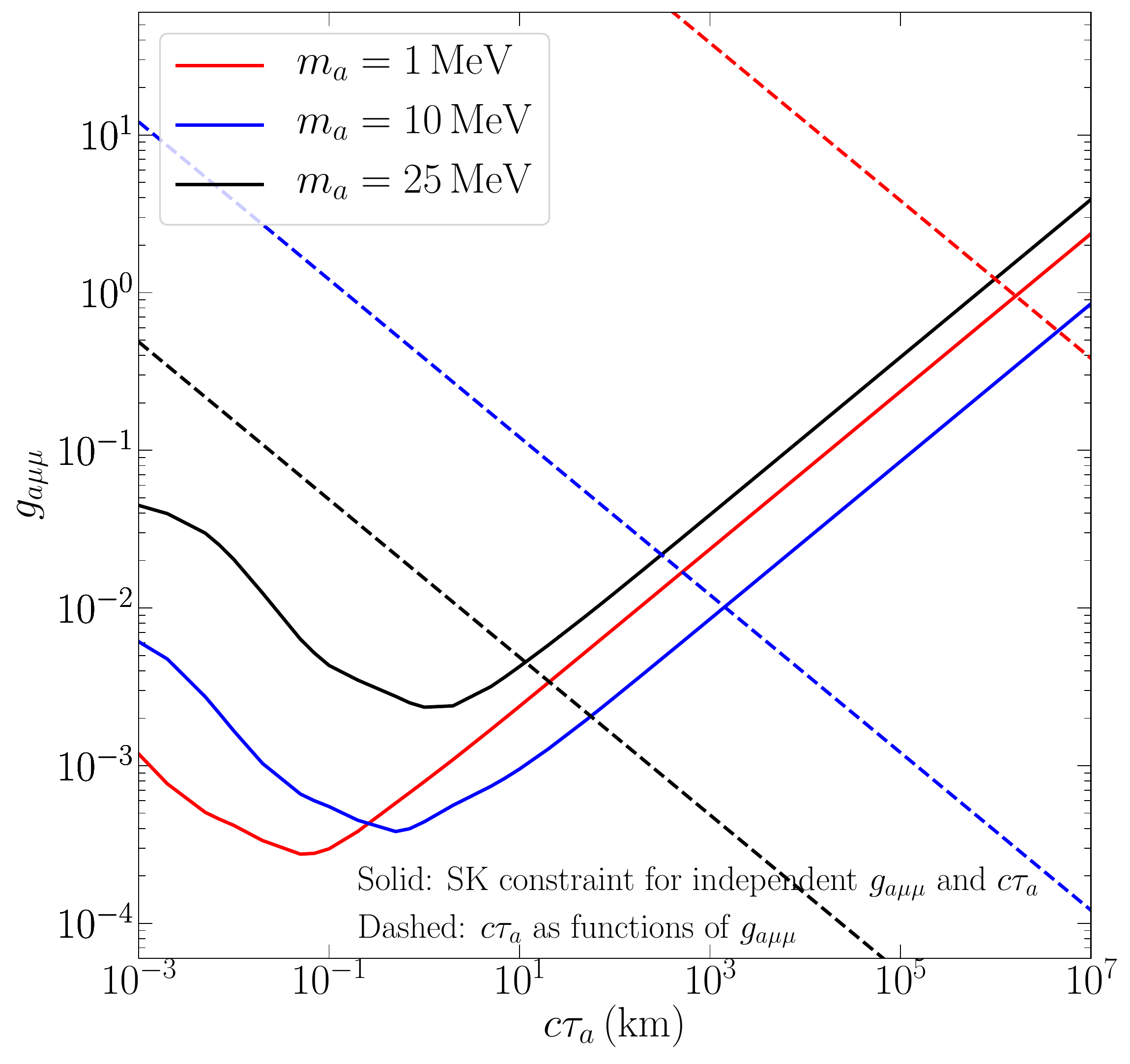}\hfill
\includegraphics[width=0.48\textwidth]{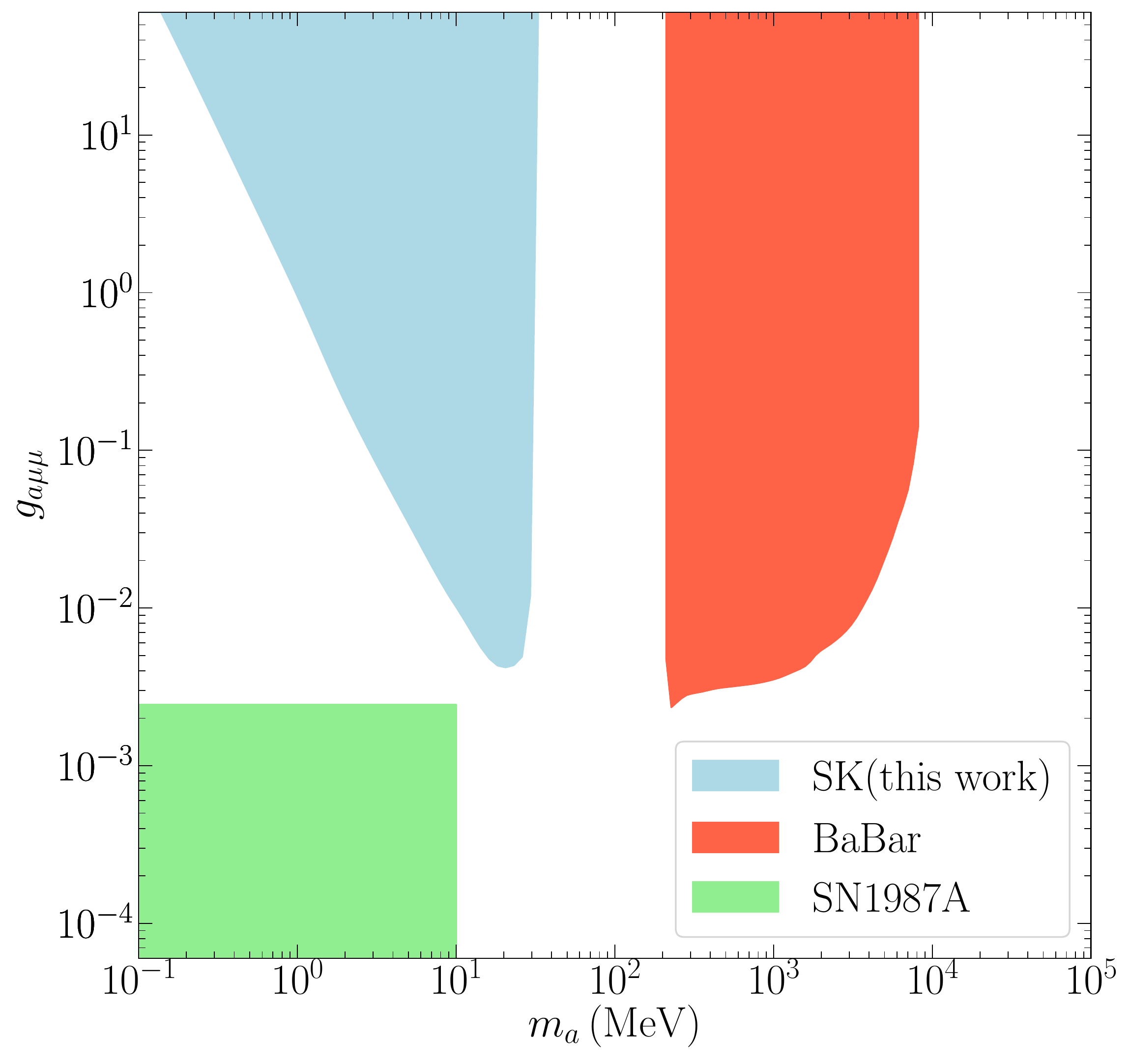}
\end{center}
\caption{\textit{Left panel}: $90\%$ C.L.~sensitivity reach of SK to the muonphilic ALPs for independent $g_{a\mu\mu}$ and $c\tau_a$ (solid curves), and $c\tau_a$ as a function of $g_{a\mu\mu}$ according to Eq.~\eqref{eq:ALP_lifetime} (dashed lines) in the ($c\tau_a$, $g_{a\mu\mu}$) plane, for three benchmark values of $m_a$: 1, 10, and 25 MeV.
\textit{Right panel}: constraints on ($m_a$, $g_{a\mu\mu}$) assuming $c\tau_a$ is proportional to $1/g^2_{a\mu\mu}$.
Note that $g_{a\mu\mu}$ always induces the ALP production from the charged pion decays.
For comparison, we also include the constraint from BaBar which holds only for larger $m_a$~\cite{BaBar:2016sci}, and the bounds from SN1987A which cover $g_{a\mu\mu} \sim [10^{-10}, 2\times 10^{-3}]$ for $m_a \leq 10\,{\rm MeV}$~\cite{Caputo:2021rux}.
}
\label{fig:SK_reach}
\end{figure*}

We perform $\chi^2$-fit to the SK data mentioned in Sec.~\ref{sec:SuperK} using \cite{Arguelles:2019ziu}
\begin{align}
\chi^2_i = 2 \left\lbrace N_{\rm sig}^i + N_{\rm bkg}^i - N_{\rm obs}^i \left[ 1- \log\left( \dfrac{N_{\rm obs}^i}{N_{\rm sig}^i + N_{\rm bkg}^i}\right) \right] \right\rbrace\,,
\end{align}
where $N_{\rm sig}^i$, $N_{\rm bkg}^i$, and $N_{\rm obs}^i$ are numbers of expected signal events of ALP, background events, and observed events in each data bin, respectively.
For each bin, the expected ALP signal events can be computed by
\begin{align}
N_{\rm sig}^i = \int^i dT_a d\cos\theta \dfrac{d^2 N_{\rm event}}{dT_a d\cos\theta}\,.
\end{align}
The background and observed events are extracted from Ref.~\cite{Super-Kamiokande:2017yvm}.
For a total of 10 bins considered in this work, we derive the 90\% C.L. constraint by requiring $\Delta \chi^2 \equiv \chi^2 - \chi^2_0 \leq 4.865$ with $\chi^2 = \sum_i \chi_i^2$ and $\chi^2_0$ being the case without ALP contribution.

We show the $90\%$ C.L.~sensitivity reach of SK in Fig.~\ref{fig:SK_reach} for the case where $g_{a\mu\mu}$ and $c\tau_a$ are independent of each other (solid lines in the left panel) and the case where $c\tau_a$ depends on $g_{a\mu\mu}$ (blue area in the right panel).
In the left panel, we observe that for $m_a = 1\,{\rm MeV}$, the best sensitivity of SK lies at $c\tau_a \sim 5\times 10^{-2}\,{\rm km}$, while for larger $m_a$, the best sensitivity of SK is reached at larger $c\tau_a$. 
In the same plot we also overlap the $g_{a\mu\mu}$-$c\tau_a$-independent sensitivity curves with dashed lines depicting $c\tau_a$ as a function of $g_{a\mu\mu}$ using Eq.~\eqref{eq:ALP_lifetime}.
This allows us to crosscheck with the right panel where we assume both production and decay are mediated by $g_{a\mu\mu}$; the intersection of the solid and dashed lines in the left plot for each fixed mass should coincide with the corresponding parameter point on the outer edge of the blue area in the $g_{a\mu\mu}$ vs.~$m_a$ plane in the right panel.
We find in the right plot that for $m_a = [0.1, 33]$~MeV, SK can exclude $g_{a\mu\mu} = [5\times 10^{-3}, 50]$, comparable to the BaBar exclusion limits, which, however, only hold for larger $m_a$~\cite{BaBar:2016sci}.
Below $g_{a\mu\mu} \sim 2\times 10^{-3}$ and for sub-GeV $m_a$, the parameter space is covered by the SN1987A constraint;  see Ref.~\cite{Caputo:2021rux} and earlier studies~\cite{Bollig:2020xdr,Croon:2020lrf}. Note that dedicated treatment is required to derive the SN1987A constraint for $m_a > 10\,{\rm MeV}$~\cite{Caputo:2021rux,Caputo:2022mah,Caputo:2022rca}. Thus, we only show the SN1987A constraint for $m_a \leq 10\,{\rm MeV}$ in the right panel of Fig.~\ref{fig:SK_reach}.
For each mass value within the sensitive range, the SK exclusion limits are bounded from both top and bottom.
This is because when $g_{a\mu\mu}$ is too small, the production rate of the ALPs is insufficient and the decay length is too long.
On the other hand, with a too large $g_{a\mu\mu}$, despite the enhanced production rate, the decay length is so short that the ALPs decay before reaching the SK detector.
We note that future muon beam-dump experiments can improve the sensitivity at the ALP mass range of interest in this work down to around $g_{a\mu\mu}\sim 6\times 10^{-6}$; see Ref.~\cite{Chen:2017awl} for details.
The results presented here are only based on charged pion flux.
One can in principle extend the constraint to $m_a \geq m_{\pi^\pm} - m_\mu$ by further including heavier mesons such as kaons that have a similar decay channel.
However, once $m_a \geq 2 m_\mu$ is fulfilled, the ALPs can decay into two muons, resulting in further complication during the air showers; we reserve this possibility for future work.

In addition to ALP-muon interactions, we have checked the case that the ALP has a direct coupling to the SM photons. 
In this case, the main ALP production channel in air showers is $\pi^0$ decay, with a smaller branching ratio compared to charged pion decay as a result of the much shorter lifetime of $\pi^0$. 
Therefore, we infer that given a fixed decay rate of $\pi \rightarrow a$, for $T_a < \epsilon_{\pi^\pm}$ the ALP flux from $\pi^0$ is less intense than that from $\pi^\pm$, but for higher masses the ALP flux from $\pi^0$ dominates, as $\pi^\pm$'s will not all decay well before reaching the Earth's surface.
We find that this excludes $g_{a \gamma \gamma} \gtrsim 10^{-2}\,{\rm GeV}^{-1}$ for $m_a \sim \mathcal{O}(1\text{--}100\,{\rm MeV})$, assuming the decay of ALP is independently determined by $c\tau_a$. 
Since $g_{a \gamma \gamma} \gtrsim 10^{-2}\,{\rm GeV}^{-1}$ is already excluded by accelerator and collider constraints (see Ref.~\cite{dEnterria:2021ljz} for a recent summary), we do not demonstrate the result in this work.  
In addition, we note that in this scenario, the case that the production and the decay are both mediated by $g_{a\mu\mu}$ cannot be probed by SK, since the tree-level decay results in a much shorter $c \tau_a$ for the given couplings; see also other scenarios with tree-level decays discussed in Ref.~\cite{Arguelles:2019ziu}.

\section{Conclusions}\label{sec:conclusions}

Similar to QCD axions, axion-like-particles (ALPs), denoted as $a$ in this work, are also among the most plausible candidates of dark matter.
With their mass and couplings to the SM particles decoupled, such exotic particles are being searched for at various experimental facilities across a wide range of masses.
While the ALPs can in theory couple to various types of particles, here, we have chosen to focus on the case that the ALPs are dominantly or solely interacting with the SM muons at tree level.
In addition, we have restricted ourselves to ALP mass below the muon-pair threshold so that the ALPs only decay radiatively into a pair of photons.
Such ALPs can be produced from charged pion decays, $\pi^\pm\to \mu^\pm \nu a$ via the ALP-muon coupling $g_{a\mu\mu}$.

Large numbers of mesons including charged pions are produced in the atmospheric air showers resulting from cosmic rays.
Once ALPs are produced from these charged-pion decays, if long-lived, they can travel tens of kilometers downwards to the Earth's surface thanks to the large Lorentz boost, and decay in large-volume neutrino experiments such as Super-Kamiokande (SK), leading to Cherenkov signal events.

We make use of the numerical tool \texttt{MCEq} in order to estimate the ALP flux at the Earth's surface stemming from cosmic-ray-induced atmospheric air showers of the charged pions.
We further compute the signal event rates, taking into account the differential ALP flux, detector efficiency, data-taking time, etc., at SK.
In addition, there are background events mainly stemming from neutral pion decays into two photons as well as neutrino-induced electron-like events that lead to multiple Cherenkov rings in the electromagnetic showers.
We have extracted the level of these background events from Ref.~\cite{Super-Kamiokande:2017yvm}, and evaluated the sensitivity reach of the SK experiments to such muonphilic ALPs.
Results are presented for both cases in which the production and decay rates of the ALP are mediated by the same coupling and are decoupled, respectively, shown in Fig.~\ref{fig:SK_reach}.
In particular, we find that if both production and decay are mediated by the coupling $g_{a\mu\mu}$, SK can probe $g_{a\mu\mu}$ down to $5\times 10^{-3}$ at $m_a\sim 20$ MeV, complementary to the exclusion limits obtained at BaBar which is sensitive to larger masses, as well as SN1987A constraint which covers $g_{a\mu\mu} \lesssim 2\times 10^{-3}$.

Additionally, we have commented on further possibilities such as the inverse-Primakoff process, and the case that the ALP is coupled to the SM photons at tree level.
While the former is expected to be dominated over by the main process considered in this work, the latter is checked to give only rather weak limits that have already been excluded by past experiments.

We have also briefly discussed the implication of the anomalous muon magnetic moment measurements on our model.
Based on the current sensitivity of Fermilab $(g-2)_\mu$ measurement and once theoretical uncertainties are clarified, $|g_{a\mu\mu}|\gtrsim 5\times 10^{-4}$ can be probed for $g_{a\gamma\gamma}$ derived from $g_{a\mu\mu}$ and $m_a\le 2 m_\mu$.
However, with this coupling correlation, the ALPs always contribute negatively to $a_\mu$.
Therefore, our model is unable to alleviate the present tension between the SM prediction and observation results.

Before closing, we comment on the sensitivities of other present and future neutrino telescopes to the scenario considered in this work.
For the future Hyper-Kamiokande experiment (HK)~\cite{Abe:2011ts,Hyper-KamiokandeWorkingGroup:2013hcb}, its fiducial volume is increased by a factor 25 compared to SK.
Therefore, we naively estimate that it can improve the sensitivity reach to $g_{a\mu\mu}$ by a factor $\sqrt{5}$ ($5^{1/4}$) for the case where the production and decay of the ALP are independent (inter-dependent).
We note that the precise sensitivity reach ultimately depends on the detector configuration of HK. Moreover, if the signal discrimination rate is also enhanced, we expect the sensitivity reach could potentially enclose a large portion of the parameter space currently covered by the SN1987A constraint.
Finally, we note that in principle IceCube~\cite{IceCube:2016zyt} is also capable of probing the atmospheric ALP flux considered in this work. However, since IceCube focuses on the ultrahigh energy range, the best sensitivity lies at $c\tau_a \sim 5\times 10^{-5}\,{\rm km}$ for $m_a \sim 10\,{\rm MeV}$~\cite{Arguelles:2019ziu}; therefore, with a smaller $c\tau_a$ needed, we can expect that the constraint on $g_{a\mu\mu}$ will be much weaker, despite a much larger fiducial volume at IceCube.

\section*{Acknowledgment}
We thank Giovanna Cottin, Felix Kling,  V\'ictor Mu\~{n}oz, and Edoardo Vitagliano for useful discussions.
Z.S.W. is supported by the Ministry of Science and Technology (MoST) of Taiwan with grant number MoST-110-2811-M-007-542-MY3. J.L.K. is supported by the U.S. National Science Foundation (NSF) Theoretical Physics Program, Grant PHY-1915005.
P.Y.T. and K.C. are supported in part by MoST with grant numbers MoST-111-2112-M-007-012-MY3 and MoST-110-2112-M-007-017-MY3, respectively.

\newpage
\appendix

\begin{widetext}
\section{Energy spectrum of $\pi^\pm \to \mu^\pm \nu a $}
\label{app:energy_spectrum}
To derive the energy spectrum of $\pi^\pm(p_1) \to  \nu(p_2) +\mu^\pm (p_3)+ a(p_4) $, we start with the calculation of its amplitude $\mathcal{M}$. 
With the Lagrangian given in Eq.~\eqref{eq:ALP-muon_Lagrangian}, we can write down
\begin{align}
\sum\limits_{\rm spins} |\mathcal{M}|^2 = \dfrac{4 f_\pi^2 g_{a\mu\mu}^2 G_F^2 V_{ud}^2 \left[m_a^2 m_\mu^2 (s_{34}-m_\pi^2) + (m_\mu^2 -s_{34})(m_\mu^2 m_\pi^2 -s_{23} s_{34})\right]}{(m_\mu^2 - s_{34})^2}\,,
\end{align}
where $f_\pi = 93\,{\rm MeV}$ is the pion decay constant, $G_F$ is the Fermi constant, $V_{ud}$ is the CKM matrix element for transition from up to down quark, $s_{23} \equiv (p_2 + p_3)^2$, and $s_{34} \equiv (p_3+p_4)^2$.
We then derive the decay rate of $\pi^\pm \to \mu^\pm \nu a $, which reads
\begin{align}
\Gamma_{\pi\to\mu\nu a} = \dfrac{1}{2m_\pi} \int d\Pi_{i=2,3,4} (2\pi)^4 \delta^4(p_1 - p_2 -p_3 -p_4) \sum\limits_{\rm spins} |\mathcal{M}|^2\,,
\end{align}
where $d\Pi_i$ are Lorentz invariant phase space elements. 
Utilizing the Dalitz plot for three-body decay we then derive the differential decay rate in the rest frame of $\pi^\pm$
\begin{align}
\dfrac{d\Gamma_{\pi\to\mu\nu a}}{dE_a^*} =\dfrac{1}{128\pi^3 m_\pi^2} \int^{s_{34}^+}_{s_{34}^-} ds_{34}\,  \sum\limits_{\rm spins}|\mathcal{M}|^2\,,
\end{align}
with $E_a^*$ being the ALP energy in the rest frame of $\pi^\pm$ and the integration boundaries of $s_{34}$ given by 
\begin{align}
s_{34}^\pm = \dfrac{1}{2s_{23}} \left[ m_a^2 (s_{23} -m_\mu^2) \pm (s_{23} -m_\mu^2)\left(\sqrt{m_a^4 -2 m_a^2 (m_\pi^2 +s_{23}) +(m_\pi^2 -s_{23})^2} \mp s_{23}\right) + m_\pi^2 (m_\mu^2 +s_{23})\right]\,.
\end{align}
Note that here $s_{23} = m_\pi^2 +m_a^2 -2 m_\pi E_a^*$ and $m_a \leq E_a^* \leq (m_\pi^2 +m_a^2 -m_\mu^2)/(2m_\pi)$.
Finally, the energy spectrum of $\pi^\pm \to \mu^\pm \nu a$ can be expressed as 
\begin{align}
 \dfrac{dN_a}{dE_a^*} \equiv \dfrac{1}{\Gamma_\pi} \dfrac{d\Gamma_{\pi\to\mu\nu a}}{dE_a^*}\,,
\end{align}
where $\Gamma_\pi$ is the total decay width of $\pi^\pm$. As the branching ratio of $\pi^\pm\to\mu^\pm\nu a$ is suppressed, we adopt the SM predicted value of $\Gamma_\pi$ in this work.
Next, we apply a Lorentz transformation and average over the angular degree of freedom to obtain the ALP energy spectrum in the lab frame $dN_a/dT_a$.
The Jacobain $|\partial E_a^*/\partial T_a|$ in Eq.~\eqref{eq:energy_spectrum_lab} reads
\begin{align}
\left|\dfrac{\partial E_a^*}{\partial T_a} \right| = \dfrac{1}{m_\pi}\left[m_\pi + T_{\pi^\pm} +\dfrac{\cos\theta_a^* E_a^*  T_{\pi^\pm}(T_{\pi^\pm} +2m_\pi)}{\sqrt{T_{\pi^\pm}(T_{\pi^\pm}+2m_\pi)(E_a^* -m_a )(E_a^* + m_a)}} \right]\,,
\end{align}
where $E_a^*$ can be expressed as a function of $T_{\pi^\pm}$, $T_a$, and $\cos\theta_a^*$ with $\theta_a^*$ being the scattering angle in the rest frame of $\pi^\pm$. 

\section{Effective detection area}
\label{app:Aeff}
In this appendix, we give formulas of effective detection area, assuming the SK detector is a cylinder with radius $R_{\rm SK} = 20\,{\rm m}$ and height $H_{\rm SK} = 40\,{\rm m}$.
Gathering the effective detection areas for ALP flux coming from the top $A_1$ and from the side $A_2$, the total effective detection area reads
\begin{align}
A_{\rm eff} (T_a, \cos\theta) = |\cos\theta| A_1(T_a, \cos\theta) +|\sin\theta| A_2 (T_a, \cos\theta)\,,
\end{align}
where 
\begin{align}
A_1 (T_a, \cos\theta) &= \int_0^{R_{\rm SK}} dr \, r \int_0^{2\pi} d\phi\,\left\lbrace 1- \exp\left[ - \dfrac{\Delta l_{\rm det, 1} (r, \cos\theta ,\phi)}{c\tau_a^{\rm lab}(T_a)}\right]\right\rbrace\,, \label{eq:Aeff1} \\
A_2 (T_a, \cos\theta) &= R_{\rm SK}\int_0^{H_{\rm SK}} dh \int_{-\pi/2}^{\pi/2} d\phi\, \left\lbrace 1- \exp\left[- \dfrac{\Delta l_{\rm det, 2} (h, \cos\theta ,\phi)}{c\tau_a^{\rm lab}(T_a)} \right]\right\rbrace \,,\label{eq:Aeff2}
\end{align}
with $c\tau_a^{\rm lab}$ being the ALP decay length in the lab frame.
The ALP trajectories inside the detector are
\begin{align}
\Delta l_{\rm det, 1} (r,\cos\theta,\phi) &\equiv {\rm min} \left[ \dfrac{H_{\rm SK}}{|\cos\theta|}, \dfrac{R_{\rm SK}\sqrt{1-(r^2/R_{\rm SK}^2) \sin^2 \phi}+r\cos\phi}{|\sin\theta|}\right]\, \text{ and}\\
\Delta l_{\rm det, 2} (r,\cos\theta,\phi) &\equiv {\rm min}\left[\dfrac{H_{\rm SK}-h}{|\cos\theta}, \dfrac{2R_{\rm SK}\cos\phi}{|\sin\theta|} \right]\,,
\end{align}
for $A_1$ and $A_2$, respectively.
Note that the integration over the azimuthal angle $\phi$ is carried out in Eq.~\eqref{eq:Aeff1} and Eq.~\eqref{eq:Aeff2}.

\end{widetext}

\bibliography{refs}

\end{document}